# Spatial resolution of late reverberation in virtual acoustic environments

Christoph Kirsch[1*], Josef Poppitz[2], Torben Wendt[1,2], Steven van de Par[2], Stephan D. Ewert[1]

[1] Medizinische Physik and Cluster of Excellence Hearing4all, Carl von Ossietzky Universität Oldenburg, Germany

[2] Akustik and Cluster of Excellence Hearing4all, Carl von Ossietzky Universität Oldenburg, Germany

*Electronic mail: christoph.kirsch@uol.de

# ABSTRACT

Late reverberation involves the superposition of many sound reflections resulting in a diffuse sound field. Since the spatially resolved perception of individual diffuse reflections is impossible, simplifications can potentially be made for modelling late reverberation in room acoustics simulations with reduced spatial resolution. Such simplifications are desired for interactive, real-time virtual acoustic environments with applications in hearing research and for the evaluation of hearing supportive devices. In this context, the number and spatial arrangement of loudspeakers used for playback additionally affect spatial resolution.

The current study assessed the minimum number of spatially evenly distributed virtual late reverberation sources required to perceptually approximate spatially highly resolved isotropic and anisotropic late reverberation and to technically approximate a spherically isotropic diffuse sound field. The spatial resolution of the rendering was systematically reduced by using subsets of the loudspeakers of an 86-channel spherical loudspeaker array in an anechoic chamber. It was tested whether listeners can distinguish lower spatial resolutions for the rendering of late reverberation from the highest achievable spatial resolution in different simulated rooms. Rendering of early reflections was kept fixed. The coherence of the sound field across a pair of microphones at ear and behind-the-ear hearing device distance was assessed to separate the effects of number of virtual sources and loudspeaker array geometry. Results show that between 12 and 24 reverberation sources are required.

# INTRODUCTION

Room acoustics simulation enables the well-controllable and repeatable presentation of reverberant acoustic environments, whether they exist in reality or just in the virtual domain. While artificial reverberation has historically been used as a tool for artistic expression (Välimäki et al., 2012), room acoustics simulations are also utilized for planning of spaces (e.g., concert halls and classrooms; Rindel, 2001). Recently, room acoustics simulations and virtual acoustic environments (VAE) have gained interest as tools for psychoacoustic research and hearing aid development (Rindel, 2001; Seeber et al., 2010; Pausch et al., 2018; Grimm et al., 2018; Schutte et al., 2019; Ahrens et al., 2019; Huisman et al., 2019; Pausch and Fels, 2020), particularly in the context of ecological validity, where the results of speech-in-noise and hearing test with simplified 'synthetic' stimuli have been questioned with regard to real-world outcomes (Cord et al., 2007; Jerger, 2009; Miles et al., 2020; Keidser et al., 2020).

Loudspeaker-based rendering of VAEs enables psychoacoustic tests with free head movement as well as the operation of (hearing aid) microphone arrays within the simulated sound field. The use of VAEs for hearing research and development and evaluation of hearing devices, requires knowledge about perceptual and technical limitations of the rendered sound field: For perception and binaural hearing aid algorithms (e.g., Kollmeier et al., 1993; Dörbecker and Ernst, 1996; Srinivasan, 2008), the typical ear distance of receivers is relevant in addition to the closer spacing of multi-microphone arrangements typically used to achieve spatial directivity in behind-the-ear (BTE) hearing devices (e.g., Yousefian and Loizou, 2011; Doclo et al., 2015; Thiemann et al., 2016).

Given that acoustic communication often takes place in enclosed spaces, the simulation and rendering of late reverberation and sound reflections at boundaries of these spaces are important for VAEs. While the direct sound and distinct early reflections are typically rendered as individual sound sources (e.g., Schröder, 2011; Wendt et al., 2014), diffuse late reverberation results from a superposition of many densely spaced reflections that are spatially more or less evenly distributed. In rooms, diffuse reverberation typically dominates the after the $3^{rd}$ reflection order (Kuttruff, 1995). Depending on room geometry and the spatial distribution of sound absorption at room boundaries, the resulting diffuse late reverberation can be considered spherically isotropic or anisotropic, containing limited spatial directivity (Luizard et al., 2015; Lachenmayr et al., 2016; Romblom et al., 2016; Alary et al., 2019a). In principle, this allows for a reduced spatial resolution of the late reverberation rendering, relevant for computational efficiency in interactive, low-latency real-time VAEs. Considering hearing research and hearing devices evaluation in VAEs, the minimum required spatial resolution of late reverberation, depending on the number of (virtual) sound sources or loudspeakers, is of interest from both a perceptual and technical perspective.

Hiyama et al. (2002) found that perceptually, a specific horizontal arrangement of four loudspeakers separated by 90° can already suffice to reproduce the spatial impression of a diffuse sound field. With such a low number of loudspeakers, however, the results were strongly dependent on the rotation of the loudspeaker array with regard to the listener. In VAE applications, where listeners can freely rotate their head, such rotation dependency is problematic. Regarding the perceptual impression of diffuse sound in three-dimensional arrangements ITU-R, BS.2159-8 (2019) recommends two vertically offset rings consisting of 8 loudspeakers each. Based on informal listening tests with binaural reproduction, Laitinen and Pulkki (2009) reported a number of 12 to 20 virtual loudspeakers to be adequate for the reproduction of diffuse sound.

Grimm et al. (2015) evaluated different spatial loudspeaker reproduction methods in a simulated 2-dimensional circular loudspeakers array. It was found that 8 loudspeakers are sufficient for localization of a sound source in the horizontal plane according to an auditory model (Dietz et al., 2011) that estimates ILD and ITD. Furthermore, it was shown that a number of 8 up to 72 loudspeakers was required for specific hearing aid algorithms using microphone arrays in diffuse background noise.

Oreinos and Buchholz (2014) found similar signal-to-noise-ratio (SNR) benefits for an adaptive, binaural, correlation-based beamformer in a comparison between a real environment and an auralization via a 41 loudspeaker spherical array, driven by simulation and mixed-order ambisonics (Favrot et al., 2011) recordings. In (Oreinos and Buchholz, 2016), they additionally observed similar performance in the VAE compared to the real environment for speech intelligibility and acceptable noise level for hearing-aid equipped listeners.

Taken together, these studies provide recommendations for the rendering of diffuse sounds and demonstrate the general function of hearing aid algorithms in VAEs. However, further systematic investigation of the minimal spatial resolution required for rendering of diffuse late reverberation in the context of VAEs is required, given that: i) In contrast to earlier studies focusing on isotropic sound fields, in VAEs late reverberation occurs in conjunction with direct sound and early reflections. ii) Earlier studies often focused on cylindrically isotropic (2-dimensional) sound fields with horizontal loudspeaker arrangements (Hiyama et al., 2002; Grimm et al., 2015) in contrast to spherically (3-dimensional) isotropic or anisotropic sound fields occurring in the late reverberation. iii) Rotation of the listener in the loudspeaker array has been shown to be critical for a very low numbers of loudspeakers (Hiyama et al., 2002). iv) A connection of technical limitations of the reproduced sound field in a loudspeaker array and perception may help to guideline their design.

In the current study, spatial resolution of diffuse late reverberation was assessed in different simulated rooms by varying the number of spatially evenly distributed virtual reverberation sources (VRS) rendered in a 3-dimensional 86-channel loudspeaker array. For the perceptual evaluation two psychoacoustic experiments with normal-hearing listeners were conducted using speech and transient stimuli: 1) The first experiment used VAEs with homogenous boundary conditions, resulting in an isotropic late reverberation. 2) For the second experiment, inhomogeneous sound absorption properties were assigned to the boundaries of the virtual room in order to create an anisotropic late reverberant field. 3) In a technical evaluation, the ability to reproduce the coherence between omnidirectional receivers at typical ear and BTE microphone distance in an isotropic sound field (e.g., Cook et al., 1955) in the loudspeaker array was investigated and related to the psychoacoustic findings. Coherence at ear distance is considered relevant for psychoacoustic processes (e.g., Faller and Merimaa, 2004; Grosse et al., 2015) and has been suggested to assess the reproduction of diffuse sound fields (e.g., Hiyama et al., 2002; Walther and Faller, 2011). Limitations imposed by number of VRS and array geometry are investigated by comparison of simulated sound fields to the analytic reference coherence function for the ideal spherically isotropic sound field. Additionally, for an example test case representing a real-world environment with long reverberation time and containing direct sound and early reflections, measured binaural room impulse responses (BRIRs) and simulated BRIRs with different spatial resolution of the late reverberation were compared in terms of coherence between the microphones of a dummy head and channels of typical BTE hearing aid devices.

The method used for simulation and rendering of the VAEs is freely available at www.razrengine.com.

## Methods

### Room acoustics simulation

To vary the spatial resolution of late reverberation, the room acoustics simulator (RAZR; Wendt et al., 2014) was used. RAZR generates perceptually plausible room impulse responses (RIRs) using a geometrical acoustics-based ISM (Allen and Berkley, 1979; Borish, 1984) and a shoebox approximation of room geometry and a computationally efficient FDN (Jot and Chaigne, 1991) for late reverberation. Perceptual plausibility of the resulting room acoustics simulations was demonstrated by favorable performance in comparison to other state-of-the-art approaches in (Brinkmann et al., 2019, see their Figure 8). Early (specular) reflections up to the third reflection order were generated with the ISM. The late reverberation tail was generated by a 96-channel FDN which is fed by the last order of reflections from the ISM. The output channels of the FDN were used as virtual reverberation sources that were spatially evenly distributed around the listener (see Wendt et al., 2014 for details).

To adjust the spatial resolution of late reverberation, pairs of FDN output channels were added, resulting in "downmixes" for a reduced number of 48, 24, 12 and 6 VRS in addition to 96 VRS. Average correlation coefficients < 0.03 between the VRS signals enable the approximation of diffuse late reverberation as superposition of spatially distributed incoherent sound sources (see also Jacobsen and Roisin, 2000). By using the fixed, high number of 96 channels in the FDN, independent of the number of VRS, the spatial resolution of the late reverberation can be adjusted while maintaining the spectro-temporal characteristics and thus avoiding timbre changes in the resulting reverberant tail (see, e.g., Schlecht and Habets, 2015, 2017). Polyhedra centered around the listener were used to determine the directions for spatialization of the VRS, where the number of vertices corresponded to the number of VRS (see Figure 1). The polyhedra were directionally aligned with the (shoebox) room boundaries and were optimized for sphericity (Wadell, 1935), which ranged from 0.86 for 6 VRS to 0.99 for 96 VRS. For 6 VRS, the resulting directions are orthogonal to each other and for 12 VRS, they correspond to points lying on diagonals of a room aligned cube. Directions for VRS numbers of 24 and above were derived from a combination of 1, 2, and 4 snub cuboctahedra (snub cubes) resulting in 4, 8, and 16 VRS assigned to each of the 6 surfaces of the room. 96 VRS are assumed to be a sufficiently high number to serve as a reference condition for diffuse late reverberation.

In the case of a homogeneous distribution of acoustic absorption on all surfaces, resulting in isotropic late reverberation, all VRS radiate with the same power. To render anisotropic late reverberation in case of inhomogeneous distribution of acoustic absorption, the output of each VRS is scaled to represent the mean acoustic absorption for the solid angle coved by the VRS (see Poppitz et al., 2018 for more details). Thus, with a reduced number of VRS, the spatial sampling of anisotropic features of the late reverberation is reduced.

Discrete early reflections and the VRS were rendered using an 86-channel loudspeaker array and vector base amplitude panning (VBAP, Pulkki, 1997). VBAP is a 3-dimensional panning technique that utilizes the closest 3 loudspeakers to render virtual sound sources in between the loudspeakers. The simulated RIRs that have been rendered for the loudspeaker array are referred to as multichannel room impulse responses (MRIRs) in the following. Due to path differences between the loudspeakers participating in the rendering of a virtual sound source and the listener's ears, spectral coloration artefacts can occur when using VBAP. These were accounted for with filtering based on a statistical approach described by Laitinen et al. (2014).

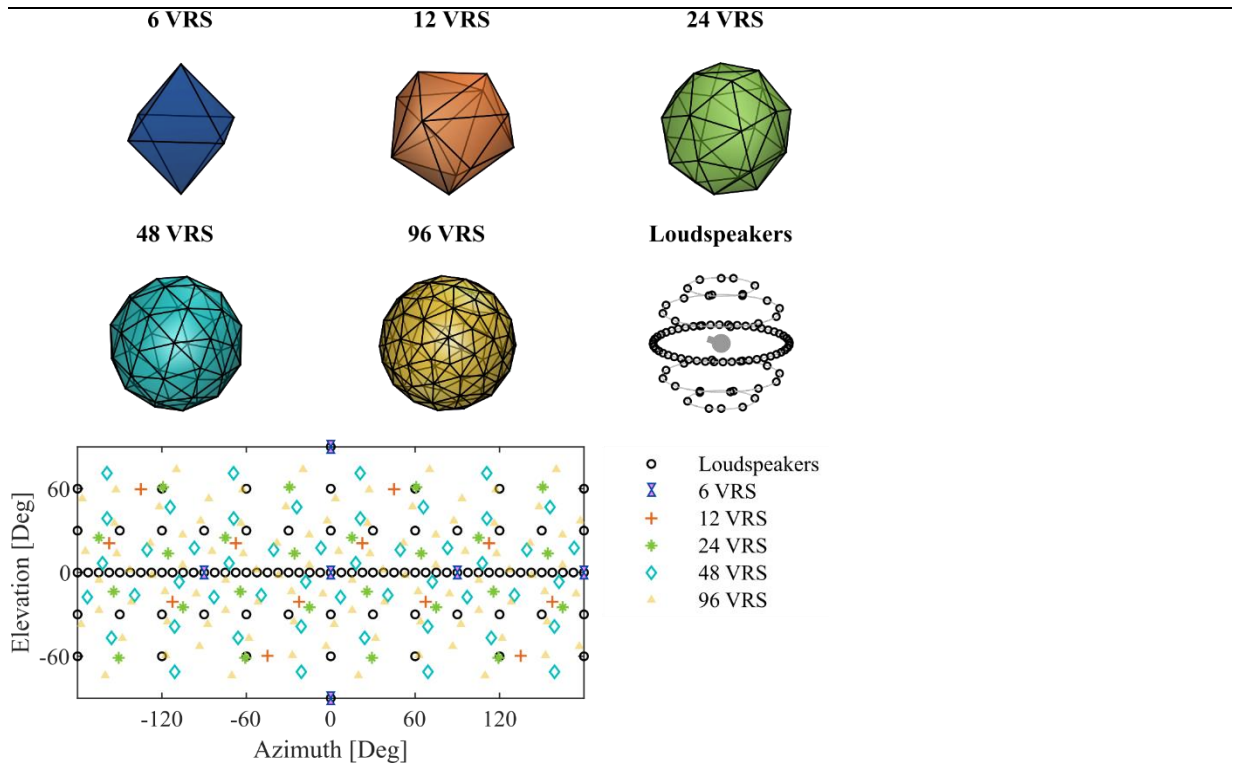

*Figure 1: Number and spatial distribution of VRS and arrangement of the 3D 86-channel loudspeaker array used for playback. The vertices of the polyhedra at the top indicate the position of the VRS. The right-hand plot in the second row shows a 3D projection of the loudspeaker positions, where the grey lines aid the visual representation of the loudspeaker arrangement in 5 rings. Polyhedra and loudspeaker array depiction are perspectively aligned. The bottom panel shows the directions of the loudspeakers (indicated as circles) and the VRS (other symbols, depending on number of VRS) in a 2D projection of the spherical coordinates relative to the default listener orientation as indicated by the nosed ball in the 3D projection of the loudspeaker positions.*

## Virtual rooms

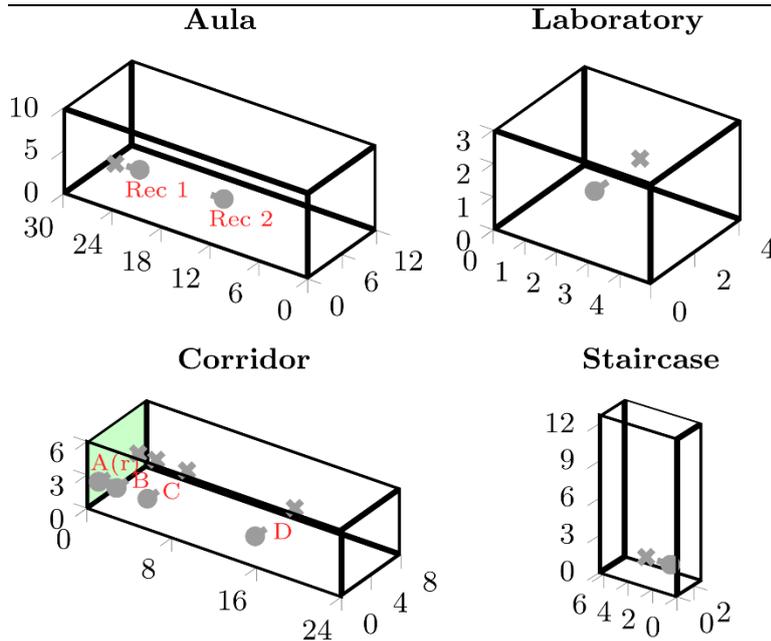

*Figure 2: Overview of the virtual rooms. Source positions are denoted by crosses, receiver positions and orientation by the nosed balls. The rooms are shown at a different scale as indicated by the dimensioning provided in m. The Aula and Laboratory (top row) were used in experiment 1, the Corridor (bottom left) was used in experiment 2, and the Staircase (bottom right) was used for the comparison with measured BRIRs. The highly absorbing surface in the corridor (bottom left) is shaded.*

A selection of four virtual, simulated rooms (denoted as Aula, Laboratory, Corridor, and Staircase) was used with multiple source-receiver combinations in some of the rooms. An overview is provided in Figure 2.

For experiment 1, two source-receiver combinations were set in a 12 m x 10 m x 30 m room (Aula, Figure 2, top-left) with a reverberation time RT60 ranging from 7.2 s at 125 Hz to 1.5 s at 8 kHz. The virtual sound source was spaced either 2.72 m from the listener position (Rec 1) or 13.01 m (Rec 2). A third condition was set in a 4.97 m x 4.12 m x 3 m room (Laboratory, Figure 2, top-right) with a RT60 of 0.4 s in the 125 Hz to 8 kHz range and a source-receiver distance of 1.7 m. In all three conditions, the receivers were oriented towards the source and all room boundaries had identical frequency-dependent absorption coefficients, resulting in isotropic late reverberation.

For experiment 2, anisotropic late reverberation was assessed using a corridor (Figure 1, bottom-left) with the dimensions 24 m x 8 m x 6 m and inhomogeneous absorption coefficients: One of the small surfaces (shaded) at the end of the corridor was highly absorbent ($\alpha = 0.99$ for all frequencies), while all other surfaces were quite reflective ($\alpha = 0.01$ to $0.11$ from 125 Hz to 8 kHz). The resulting RT60 ranged from approx. 1.3 s to 0.8 s in the frequency range from 125 Hz to 8 kHz. Several source-receiver-position combinations were considered to vary the solid angle (or field of view, FOV) occupied by the highly absorbing wall and thus the spatial features of the resulting anisotropic late reverberation. Source and receiver were always aligned on an axis parallel to the highly absorbent surface, and were located at a height of 1.8 m above the floor and 1.33 m from the sidewalls, resulting in a fixed source-receiver distance of 5.33 m. There were four source-receiver combinations at different distances h to the highly absorbing wall, so that a wide range of solid angles with a horizontal FOV γ of the absorbent wall is obtained. The exact distances and FOV angles are denoted in Tab. I. The receiver was always oriented towards the sound source (azimuth angle β was 0°) for all combinations (denoted A to D, see also Figure 2) except for Ar, where the receiver was rotated by 60° towards the direction of the corridor.

As part of the technical evaluation, a highly reverberant room (Staircase, Figure 1, bottom-right) with dimensions 2.98 m x 6.83 m x 12.71 m was used, where most of the surfaces are painted concrete. In this room, BRIRs have been measured (see Apparatus, II.3) to obtain realistic reference coherence function estimates. The source and receiver were located at the ground floor at a height of 1.72 m, spaced 2 m apart. The measured and simulated reverberation time in the staircase ranged from 5.3 s at 125 Hz to 2.6 s at 8 kHz.

*Table I: Source-receiver geometry in the Corridor. h denotes the distance of the source receiver pair from the highly absorbing surface (see Figure 1), γ describes the FOV angle and β corresponds to the azimuthal listener orientation in relation to the sound source with positive angles indicating a clockwise rotation.*

| Position | h [m] | γ [Deg] | β [Deg] |
|---|---|---|---|
| A | 0.18 | 170 | 0 |
| Ar | 0.18 | 170 | 60 |
| B | 1.87 | 110 | 0 |
| C | 4.77 | 70 | 0 |
| D | 14.93 | 30 | 0 |

**Listeners**

For experiment 1, 9 self-reported normal hearing listeners (2 female, 7 male) aged between 26 and 32 were recruited. For experiment 2, 10 listeners (5 female, 5 male) were recruited amongst students at the University of Oldenburg. They were between 21 and 28 years old at the time of testing. All were tested for normal hearing by means of pure tone audiometry (Hearing threshold ≤ 25 dBHL between 125 Hz and 8 kHz). All subjects reported varying amounts of experience with previous listening tests.

**Stimuli**

Three types of stimuli were used: i) A deterministic transient stimulus (referred to as pulse or pink pulse) was generated as digital delta pulse and filtered to achieve a pink spectrum (sampling rate 44.1 kHz, decay from 0 dBFS to -60 dBFS in 36 ms). This transient signal provides the listener with an opportunity to listen to the decay of the reverberation tail and is well suited for subjectively rating room acoustic qualities of a space (similar to using hand clapping). Given its deterministic nature, it is assumed that the pink pulse offers a high sensitivity to any change in the rendering of the RIRs. ii) A semi-transient pink noise burst with duration of 50 ms. The bursts were generated from white Gaussian noise and filtered to achieve a pink spectrum after cutting in the time domain. A different noise token was generated for each burst to avoid identical (deterministic) source signals. As for the pink pulse, the relatively short pink burst resulted in a clearly audible decay, however, with random variation in the coloration, minimizing the availability of coloration cues for the discrimination task. iii) A continuous speech stimulus composed of sentences from the OLSA corpus with different talkers (Kuehnel et al., 1999; Hochmuth et al., 2015). Different sentences were randomly selected to avoid deterministic stimuli and the availability of coloration cues, leaving subjects with only spatial cues. This condition is referred to as speech (mixed).

All stimuli were convolved with MRIRs for a specific source-receiver combination in a given virtual room. Five versions were generated deviating in the number of VRS (6 to 96), but being identical in terms of early reflections:

In experiment 1, the pink pulse and speech stimulus were used. The reverberated pulse resulted in signals of about 2000 ms (including a 50 ms fadeout) and 860 ms in duration for the Aula and Laboratory room respectively. For the Laboratory, a pause interval of 500 ms was added between consecutive stimuli. For the speech (mixed) stimulus, different sentences were randomly selected out of a subset of 6 sentences from two different talkers (1 male, 1 female). The overall duration of the stimuli was different for each of the sentences and ranged from about 3700 ms to 4500 ms including 2000 ms of the reverberant tail before being faded out. No extra silence interval separated the presentation of consecutive stimuli.

In experiment 2, the pink pulse, the pink (noise) bursts, and the speech stimulus were used. For the reverberated transient stimuli the signal duration was restricted to 1200 ms including a 100 ms fadeout and followed by a 300 ms silence (pause) interval. For the speech (mixed) stimulus, four different talkers (2 male, 2 female) were used and different sentences were randomly selected out of a set of 100 sentences from each talker. In addition to the mixed condition, a single sentence was used repeatedly referred to as speech (identical). Circular convolution was applied in order to mimic reverberation from previous sentences during an ongoing conversation. A silence interval of 100 ms was added before and after each sentence prior to circular convolution with the MRIR. After convolution, a fade in and fade out over the duration of the silence interval was applied. The overall duration of the stimuli was different for each of the sentences and ranged from approx. 1700 ms to 3000 ms. 300 ms of silence (pause) separated the presentation of consecutive stimuli.

For the pulse and burst conditions, the A-weighted, impulse-weighted sound pressure level at the listening position over the duration of 4500 ms containing three stimuli (followed by decay and silence) was approx. 65 dBA,I SPL. For the speech conditions, the equivalent A-weighted sound pressure level at the listening position over the duration of a presentation of three sentences (about 7500 ms) was approx. 65 dBAeq SPL. All signals were digitally generated and processed at a sampling rate of 44.1 kHz.

**Apparatus and procedure**

The listeners were seated on a fixed (non-rotating) chair in the center of an 86-channel 3D loudspeaker array (Genelec 8030 c/b) mounted in a 7 m x 9 m x 7 m anechoic chamber with 0.75 m foam wedges. The loudspeaker array (see Figure 1) is approximately spherical with a radius of the main (horizontal) loudspeaker ring of about 2.5 m. The loudspeakers are inhomogeneously arranged in five rings at -60°, -30°, 0°, 30°, 60° elevation and two additional loudspeakers below and above the center point (-90°, 90° elevation). The azimuthal spacing of loudspeakers is 7.5° degrees in the horizontal ring and 30° and 60° degrees respectively in the rings outside of the horizontal plane.

A computer monitor was placed straight ahead of the listeners in a distance of 2.5 m in order to inform them about the progress of the experiment and provide a direction to point their gaze at. The listeners head movement was neither constrained nor monitored, enabling natural head movements. Participants used a wireless keyboard on their lap so that responses could be given without looking at the controls.

For the listening tests an ABX paradigm was used. A reference rendering with maximum number of VRS (96) and the rendering under test (with lower number of VRS) were presented randomly as either A and B of the sequence. X was randomly chosen to be either the reference or the rendering under test. Test subjects had to determine, whether X was perceived to be similar to A or B in terms of spatial properties. Depending on the condition, the source material (prior to application of the MRIR) was not necessarily the same for A, B, and X: For the pink pulse , source material was identical throughout the entire measurement and for speech (identical) throughout each trial, while it differed for pink burst and speech (mixed).

In both experiments, the procedure was separated into runs that covered a particular combination of room condition and stimulus. Subjects were given the opportunity to take short breaks in between runs. Experiment 1 and 2 consisted of 6 and 9 runs, respectively. 20 presentations per number of VRS resulted in 80 ABX trials per run. Only exception to this was the speech (mixed) condition in experiment 2, where only 6 VRS vs. the reference have been tested, resulting in 20 ABX trials per run. One experimental run took about 10 minutes (pulse, burst) or 15 minutes (speech) to complete. Subjects were not provided feedback on the correctness of their responses. Additionally, audiometry prior to the listening test and a familiarization phase were added for experiment 2. The latter was comprised of the presentation of four ABX trials for a pink pulse, a pink burst with 6 and 12 VRS, and four trials with mixed speech and 6 VRS. During the familiarization phase, subjects received feedback on the correctness of their responses. For experiment 2, the experimental procedure, except for the audiometry, was repeated in an additional session on a different day resulting in overall 40 representations per VRS number.

**Technical evaluation**

The focus of the technical evaluation was on the ability to reproduce i) interaural (head-size spaced) signal properties relevant for binaural perception and binaural hearing aid algorithms, and additionally on ii) closely spaced inter-microphone signal properties relevant for multi-microphone hearing aid signal processing in an approximated spherically isotropic sound field given the number and positions of the VRS. The technical evaluation focused on the spherically isotropic case for which a clearly defined target coherence function (reference) exists.

Coherence was used as technical measure to assess the quality of the reproduction of a diffuse sound field, generated by independent Gaussian noises as output of the VRS. All frequency-dependent coherence estimates Cxy(f) between the signals x and y have been calculated according to:

$$C_{xy}(f) = \Re\left(\frac{G_{xy}(f)}{\sqrt{G_{xx}(f)G_{yy}(f)}}\right) \quad (1)$$

where *f* denotes the frequency, $\Re$ is the real-part operator and *G* represents the spectral density estimate according to Welch. The calculations were performed for consecutive windows with an 75 % overlap and a length of 512 samples at 96 kHz sampling rate to obtain an average coherence estimate. The Gaussian noise VRS signals were 30 s in duration.

The coherence was estimated for omnidirectional receivers. The (reference) coherence Γ between two omnidirectional receivers spaced with the distance d in a spherically isotropic sound field can, according to, e.g., Cook et al. (1955), be described as:

$$\Gamma_{sph}(\omega) = \text{sinc}\left(\frac{\omega d}{c}\right) \quad (2)$$

where ω is the circular frequency and c is the speed of sound.

In the simulation, two omnidirectional receivers were symmetrically positioned near the center of the array on an axis orthogonal to the 0° Azimuth and Elevation direction as defined in Figure 1. The spacing between the receivers for the simulations was 170 mm, approximating ear distance and the width of a human head, and 15.6 mm, corresponding to the distance between the front and rear microphones of the BTE hearing aids used in the dummy head recordings of the Staircase room.

Idealized IRs have been derived for the paths from sound sources to the receivers for three arrangements: i) For the current 86-channel array to the receivers, ii) for a different array geometry with a similar number of loudspeakers arranged in a homogenous spacing across the spherical surface according to a Fibonacci lattice (González, 2009), and iii) for the VRS positions, representing a direct spatialization with loudspeakers at the VRS positions. In all three cases, a sphere diameter of 2.5 m was assumed. The effect of air absorption was neglected, meaning that each idealized IR consisted of a delay and attenuation resulting from the distance between the sound source and the receiver. Receiver signals have been generated by convolving independent Gaussian noises with the IRs and summing the resulting signals for both receivers.

In addition to the systematic coherence assessment for the spherically isotropic sound field, the Staircase room served as an example scenario to assess coherence in the VAE with simulated BRIRs using different numbers of VRS in comparison to the measured BRIRs in a real room. To generate the simulated BRIRs, a set of head-related impulse responses (HRIRs) has been measured for all loudspeakers in the 86-channel array (sampling rate 44.1 kHz) using a G.R.A.S. KEMAR type 45BM head and torso simulator equipped with behind-the-ear (BTE) hearing aid microphone dummies (see Kayser et al., 2009). Each hearing aid dummy had three microphones, resulting in a total of 8 measured channels (eardrum, BTE front, mid and rear, for each the left and right side of the head). For each loudspeaker in the array, the respective channel of the simulated MRIR was convolved with the according HRIRs for the two eardrum microphones and the BTE front and rear channels (right side) and summed over all loudspeakers resulting in simulated BRIRs. These simulated BRIRs allow for a direct comparison of the rendering in the loudspeaker array to the according real-world BRIRs which were measured using the same equipment on the ground floor of the Staircase room (see Figure 2, bottom right).

## Perceptual evaluation

### Experiment 1: Isotropic late reverberation

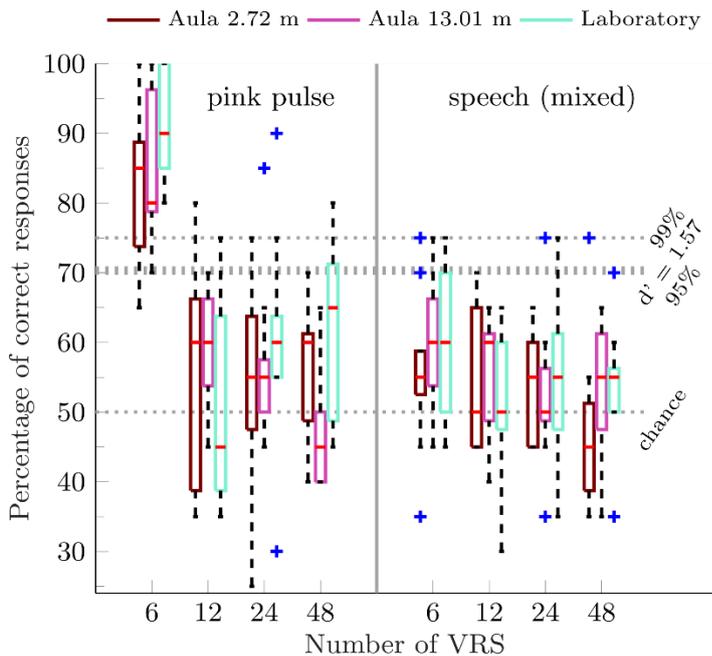

*Figure 3: Percentage of correct responses for conditions with isotropic late reverberation. Boxes represent the 25% to 75% percentile range and the horizontal red lines within the boxes represent the median. Whiskers denote highest or lowest value within 1.5 times the interquartile range from the box edge. Any values outside of this range were classified as outliers, represented by a blue cross. The different virtual room conditions are indicated on the top and by different colors of the boxes. Along the x-axis, the number of VRS is denoted. The three horizontal lines indicate the chance level and the 95% and 99% confidence range of a binomial distribution with p =0.5. The d'-value of 1.57 almost overlaps with the 95% confidence range here.*

Figure 3 shows the percentage of correct responses (ability to discriminate the auralization with reduced number of VRS from the reference) as a function of VRS for the pink pulse stimuli in the left section and for the speech (mixed) stimuli in the right section. The boxplots show the median and interquartile ranges across listeners for all three room conditions (Aula Rec 1, Aula Rec 2, Laboratory from left to right), grouped by the number of VRS used for rendering. Crosses indicate outliers, defined as being more than 1.5 interquartile ranges outside of the actual interquartile range. The three dashed horizontal lines correspond to chance level (50% correct) and to the percentage of correct responses required to reject the hypothesis, that responses are drawn from a 0.5 chance binomial random variable at a confidence level of 95% or 99%, as indicated on the right-hand side. These lines serve as an orientation to estimate whether the respective scores are achieved by guessing. Additionally, to assess discriminability, the percentage correct (70.7%) according to a d'-value of 1.57 for the current ABX/BAX experiment is indicated according to Macmillan and Creelman (2005; Table A5.3.) based on the assumption of subjects employing an independent observations strategy. If a differencing strategy is assumed as argued by Hautus and Meng (2002), the d'-value increases to 1.76.

For the pink pulse and 6 VRS (left-hand side), there is a high proportion of correct responses regardless of the virtual room configuration (median >= 80%). For higher numbers of VRS and the pink pulse, the median amount of correct responses is close to chance (50% correct), even though there are outliers with up to 90% of correct responses. There are no obvious tendencies with regard to the effect of the virtual room or the further increase in the number of VRS. For the speech (mixed) condition (right-hand section in Figure 3), even for 6 VRS, the median correct responses range from only 55 % to 60 % and are otherwise also close to chance as observed for the pink pulse.

A three-way repeated measures analysis of variance (ANOVA) showed a significant main effect of the number of VRS [$F(3, 24) = 42.4, p < 0.001$] and of the stimuli [$F(1,8) = 22.4, p < 0.01$], but no significant main effect of the room configuration [$F(2, 16) = 1.5, p = 0.26$]. A significant interaction was found only for the number of VRS and the stimuli [$F(3, 24) = 13.2, p < 0.01$],

Post-hoc pair-wise comparisons (Bonferroni) revealed that the main effect of the number of VRS and the interaction of VRS and stimulus can be attributed to significant differences ($p < 0.001$) between 6 VRS and all other VRS numbers for the pink pulse, while there are no significant differences in detection performance between the higher VRS numbers (12-48) for the pink pulse and no significant differences for any number of VRS for speech (mixed).

In summary, for isotropic late reverberation 12 VRS are sufficient to perceptually approximate a spherically isotropic sound field with the transient pink pulse stimulus. For speech with additional uncertainty regarding the source signal, even 6 VRS appear to suffice.

**Experiment 2: Anisotropic late reverberation**

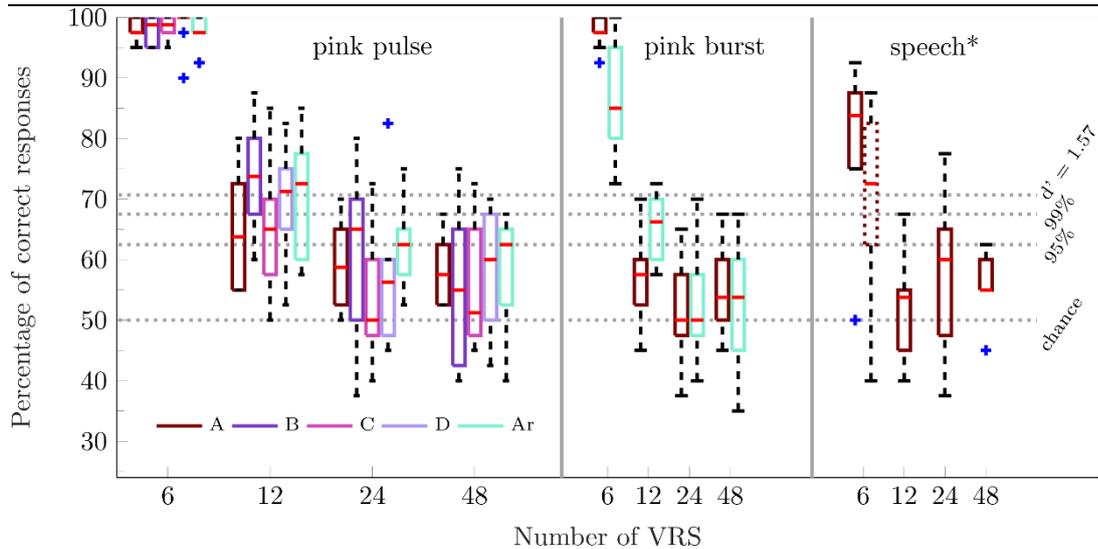

*Figure 4: Percentage of correct responses for the different experimental conditions with anisotropic late reverberation, in the same style as in Figure 3. The order of the position inside the virtual corridor is indicated at the top and by the different shades of colors of the boxes. Along the x-axis, the number of VRS is denoted. Results are grouped by stimulus. For speech (right-hand subpanel) the dotted box shows the data measured with random speech (mixed).*

Figure 4 shows the average discrimination results between conditions with fewer VRS as indicated on the x-axis and the reference condition with 96 VRS in a similar format as in Figure 3. The order of the boxes and colors indicate the room conditions with different distance from the absorbing wall (left to right: A-D, close to far; Ar, rotated) as indicated in the legend. The figure is separated into three sections by vertical grey lines for pink pulse, pink burst, and speech (identical, mixed) from left to right, grouped by the number of VRS. For pink burst and speech, only subsets of the conditions have been measured: Room conditions A and Ar for the pink burst and room condition A for the speech. For speech mixed (dotted box), only renderings with 6 VRS were compared to the reference.

A fairly narrow distribution of subject performance can be observed with 6 VRS for the pink pulse (left-hand section) on all positions and for the non-rotated (A) pink burst (center section). For these stimuli, all subjects responded correctly in at least 90% of the trials. For 12 VRS with the pink pulse, there is a wide spread at all different positions with individual performances ranging between 50 % and 87.5 % correct responses. This generally wide performance range can also be observed for 24 and 48 VRS, but with lower medians compared to 12 VRS. A consistent dependency on the position within the room (A-D) is not immediately apparent.

For the pink burst (center section), a much wider performance range with a lower median can be observed for 6 VRS for the rotated condition (Ar) compared to the non-rotated condition (A). While the performance median for the rotated condition is lower for 6 VRS when compared to no rotation, it is higher relative to the non-rotated condition for 12 VRS. Similar to the pink pulse, with the pink burst, the medians with 24 and 48 VRS are slightly lower compared to 12 VRS.

Opposed to the other source signals, speech-based stimuli (right-hand section) show a wide performance range even for 6 VRS. This is especially apparent for the mixed speech stimuli (dotted box) where individual discrimination performance ranged from 40% to 90 %. Here, subjects had to purely rely on spatial changes in the rendering. 12 to 48 VRS show performance medians at and somewhat above 50% with variability strongly reduced for 48 VRS, where none of the subjects exceed 65 % of correct responses.

Given that not all room conditions where measured with every stimulus, for statistical analysis three separate ANOVAs were performed[1]:

For the pink pulse, the effect of VRS and position within the room was assessed with a two-way repeated-measures ANOVA. A significant main effect was found for the number of VRS [$F(3, 27) = 250.26$, $p < 0.001$]. No significant main effect was found for the position within the room [$F(4, 36) = 0.84$, $p = 0.51$], and no significant interaction was found [$F(12,108) = 0.96$, $p = 0.46$]. Post-hoc pair-wise comparisons (Bonferroni) revealed significant differences between the conditions with 6 VRS compared to all other conditions. While for 12 VRS, there were some significant differences compared to 24 VRS at position C and to 48 VRS at position B and D, no significant difference was found between 24 and 48 VRS.

For the pink pulse and burst, a three-way repeated-measures ANOVA was performed to assess the effect of the number of VRS, stimulus type (pulse vs burst) and rotation (position A vs Ar). A significant main effect was found for the number of virtual sources [$F(3, 27) = 335.67$, $p < 0.001$)] and for stimulus type [$F(1,9) = 28.80$, $p < 0.001$]. No main effect was found for rotation [$F(1,9) = 12.66$, $p = 0.34$]. No significant interactions were found for the number of VRS and stimuli, the stimuli and rotation and the number of VRS, stimuli and rotation ($p > 0.2$). However, there was an interaction between the number of VRS and rotation [$F(3,27) = 3.42$, $p = 0.03$].

To additionally assess the effect of stimulus type (pulse, burst, speech) a two-way repeated-measured ANOVA was performed for the number of VRS and the different stimuli with all the results obtained at position A except for the additional measurement of speech (mixed). A significant main effect was found for the number of virtual sources [$F(3, 27) = 137.04$, $p < 0.001$] and for the stimulus type [($F(2, 18) = 12.29$, $p < 0.001$)], as well as a significant interaction [$F(6, 54) = 5.12$, $p < 0.001$]. Post-hoc pair-wise comparisons (Bonferroni) revealed a significant difference between the pink pulse and the other two stimuli, but no significant difference between the pink burst and speech (identical) condition.

Finally, a paired-samples t-test between the results obtained with 6 VRS for speech (identical)- and speech (mixed)-stimuli showed a significant mean difference of 11.75 percentage points in score, with a standard error of 2.79, $p = 0.0023$.

Taken together, the results indicate that for the pink pulse stimulus 12 to 24 source appear sufficient to render the late reverberation. If uncertainty is introduced in a transient stimulus (pink burst), already 12 sources are sufficient for the closest distance to the wall (A). Similar results are observed for the speech (identical) stimulus. A number of 6 VRS is distinguishable for all conditions, however, for speech (mixed) a clear decline of discrimination performance was observed, with some individuals even performing at chance level.

---

[1] A Gaussian distribution of the individual scores per experiment condition can only be assumed for the majority of the data according to Shapiro-Wilk testing, given that the results for 6 VRS are at the upper range of the measured scale. Sphericity was (Mauchly's test) was not violated.

**Discussion**

*Isotropic late reverberation*. For isotropic late reverberation (experiment 1) the current finding of 12 VRS as minimum number for rendering late reverberation support the default choice of 12 VRS in the original RAZR implementation (Wendt et al., 2014). The result is also generally in line with earlier findings in the literature regarding the minimum number of loudspeakers required to achieve a perceptually diffuse sound field: Laitinen and Pulkki (2009) suggested 12 to 20 loudspeakers as sufficient for a 3-dimensional arrangement. In Hiyama et al. (2002) for 8 or more horizontally arranged loudspeakers, the contribution of additional loudspeakers was strongly decreased. The current results pointing to 12 VRS were obtained for the pink pulse stimulus, which was determined to be the most critical to show effects of the number of VRS in preliminary experiments. For this transient and deterministic stimulus, subtle differences in the reverberant decay process including spatial features but also spectral coloration changes that might occur as an artefact of VBAP are easily revealed. With the random speech (mixed) stimulus, median percentage-correct values of only 55-60 % even for 6 VRS were observed, and again no difference between 12 to 96 VRS with median values around 50-55 %. Thus, depending on the stimulus and with a fixed relation between receiver orientation and VRS, even less than 12 VRS might be sufficient for spherical (3-dimensional) isotropic cases.

Furthermore, it can be assumed that cylindrical (2-dimensional) isotropic sound fields, as occurring for late reverberation with reflections resulting mainly from vertical structures (e.g., in the case of a highly absorbent or absent ceiling) are less demanding and require at least 8 horizontal loudspeakers extrapolating from 12 for the spherical isotropic case. This coincides with the findings of (Grimm et al., 2015). However, it is conceivable that the anisotropic diffuse cases as assessed in experiment 2 are more demanding, and that the number of VRS also affects the binaural parameters as the interaural coherence depending on the rotation of the listener in the loudspeaker array as discussed below in the context of experiment 2 and the technical evaluation.

*Anisotropic late reverberation*. For anisotropic late reverberation (experiment 2), reflecting a spatially more challenging condition, 6 VRS could be clearly distinguished from the reference for all stimuli, with the exception of the speech (mixed)-stimulus. Here, different (random) sentences were presented for each signal in the ABX trial so that listeners had to rely solely on spatial cues. Similar to the isotropic case in experiment 1, uncertainties in the (dry) source signal resulted in considerable differences in individual discrimination performance. The results indicate that for unknown source material, or stimuli like running speech, demands for spatial resolution of late reverberation in VAEs can be lowered even for anisotropic late reverberation. Already a comparatively small number of VRS could result in a convincing rendering for the majority of listeners, which might be useful in applications where lowest computational requirements are key (e.g., spatial rendering in teleconference systems or hearing supportive algorithms in mobile applications).

A dependency of the number of required VRS for a perceptually plausible rendering on the source material is suggested by the significant differences between the deterministic pink pulse and the other stimuli for 12 VRS and 24 VRS. As stated earlier, the deterministic pink pulse allows for a detailed comparison of spectral coloration differences in the decay process. For the random pink burst, a general decrease in detection performance is observed in non-rotated conditions with more than 6 VRS compared to the pink pulse. The stochastic nature of the pink burst stimulus and its variability throughout the ABX trials likely reduced the availability of spectral artefacts as a discrimination cue.

*Receiver rotation.* The effect of receiver orientation (A, Ar) in relation to the virtual source was assessed in experiment 2 for the pink pulse and burst stimuli. Based on the statistically significant interaction between the number of VRS and the rotation (regardless of the non-significant main effect of receiver orientation), visual assessment of the distribution of responses allows for some speculation: The most obvious deviation occurs for 6 VRS with the pink burst, where the rotated condition shows a considerably reduced median and a much wider spread in the distribution of responses, indicating that some listeners had difficulties to distinguish renderings in the rotated condition. The rotated condition differs from the un-rotated condition by a 60° azimuth incident angle of the virtual sound source and a listener rotated more towards the reverberant corridor. Considering the 6 VRS, the un-rotated condition results in a considerably larger interaural coherence compared to 96 VRS, given that 4 VRS are in the sagittal plane and produce no interaural differences and one VRS is to the right of the listener. The VRS to the left is largely attenuated, representing the highly absorbent wall in the spatially subsampled rendering. For any higher number of VRS, the interaural coherence is reduced, as more VRS outside the sagittal plane introduce interaural differences. Likewise, for the rotated condition, two VRS of the overall five active VRS are moved out of the sagittal plane, reducing interaural coherence. It is likely that the interaural coherence changes for the rotated condition, explaining the reduced detection rates. A similar effect of rotation was observed in Hiyama et al. (2002) for rendering a diffuse sound field with only four horizontally arranged speakers. In their study, the 45° rotated array with all loudspeakers introducing interaural differences was preferred over the (un-rotated) arrangement with two loudspeakers (front, back) in the sagittal plane. The slight increase in detection likelihood for 12 VRS with the rotated condition with pink bursts and the absence of such an increase for 24 VRS and 48 VRS might indicate that 12 VRS overall just barely meet the requirements and a higher number of VRS would be beneficial in some applications.

The effect of rotation on interaural coherence is also assessed in the following technical evaluation for the isotropic case.

# Technical Evaluation

## Isotropic sound field reproduction

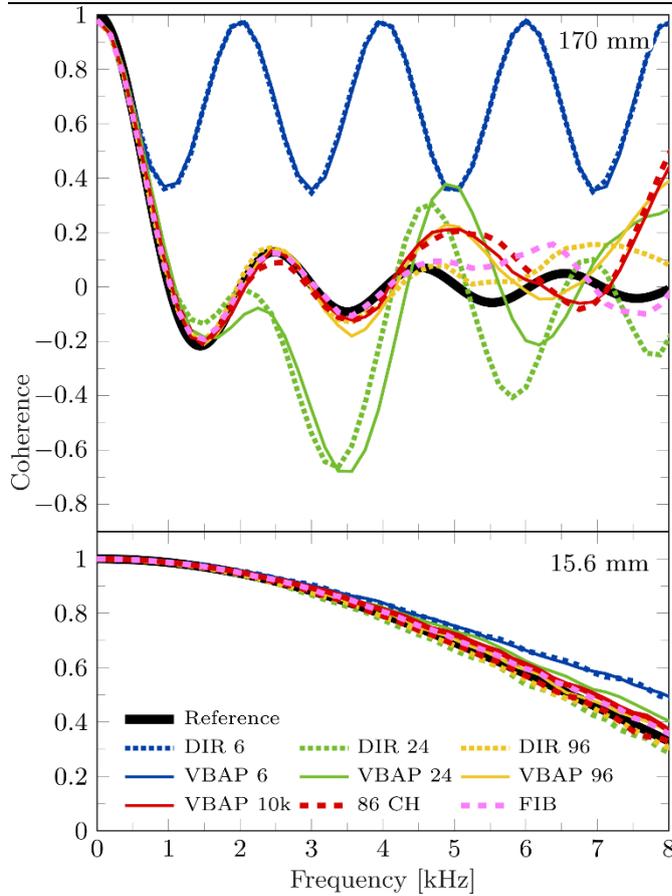

*Figure 5: Coherence function estimates between two omnidirectional receivers at 170 mm (top panel) and 15.6 mm distance (bottom panel). The reference function for a spherically isotropic sound field is shown as a thick solid line. The coherence functions for the 86-channel loudspeaker array (86 CH, dashed) with weighted incoherent loudspeaker signals, for 87 incoherent sources distributed as a Fibonacci lattice (FIB, dashed) are presented. DIR: direct spatialization of the VRS (dotted traces), VBAP: spatialization of the VRS via VBAP in the 86-channel loudspeaker array (solid traces). VBAP 10k refers to 9999 VRS.*

The top panel of Figure 5 shows coherence function estimates for two omnidirectional receivers spaced at 170 mm (ear distance) in the frequency range up to 8 kHz. The analytically derived reference function for a spherically isotropic sound field is represented by a thick black solid line. To separate the effect of loudspeaker array topology and VRS, the closest approximation possible with the specific 86-channel loudspeaker array is demonstrated with VBAP 10 k (solid red), representing 9999 VRS spatialized via VBAP, and with the condition 86 CH (dashed red). For 86 CH, the relative power of the loudspeaker channels has been derived from the VBAP case and the signals have been replaced with incoherent noise of the same power. Both VBAP 10 k and CH 86 are very similar and approximate the reference function well up to about 4.3 kHz. For higher frequencies, deviations occur with a peak in coherence at about 4.7 kHz, a minimum at about 7.2 kHz, and a steep rise towards the upper end of the displayed frequency range. These deviations from the reference are caused by the finite number of loudspeakers as well as by the inhomogeneous distribution of the loudspeakers on a sphere and will affect all VRS renderings. In comparison, the coherence function estimate for 87 loudspeakers distributed on a Fibonacci lattice with each loudspeaker emitting incoherent signals with the same power (FIB, dashed pink) shows the effect of a spatially homogeneous loudspeaker distribution on a sphere. Here, the reference is similarly matched up to 4.3 kHz while overall smaller deviations occur for higher frequencies.

Second, to separate the effect of number of VRS and VBAP, the approximation of the isotropic reference sound field was assessed as a function of the number of VRS with and without VBAP-based spatialization in the loudspeaker array. Coherence function estimates for 6, 24, and 96 incoherent VRS without VBAP (referred to as DIR) are shown as dotted lines in Figure 5. These functions would be achieved if physical loudspeakers were present at the positions of the VRS (see Figure 2) or in the case of binaural rendering to headphones using HRTFs for the respective VRS positions. The solid lines in Fig 5 are spatialized via VBAP where multiple loudspeakers typically represent a single VRS thus introducing partial coherence of the loudspeaker signals. For 6 VRS, the VRS positions match those of the loudspeakers, resulting in identical coherence function estimates for VBAP and DIR. For the 170 mm spacing in the top panel of Fig 5, a periodic behavior across the frequency range is observed, beginning to diverge from the reference at about 700 Hz. For 24 VRS, the VRS positions do not match those of the loudspeakers and a subset of 62 loudspeakers is used. Here, VBAP and DIR coherence function estimates follow the reference up to a frequency of about 1.7 kHz and exhibit increasing divergence at higher frequencies. With 96 VRS, which served as a reference condition in the perceptual evaluation, the number of VRS exceeds the number of physical loudspeakers present in the array. Due to the inhomogeneous distribution of loudspeakers in the array, only 82 of the 86 loudspeakers are used with VBAP in this condition. Coherence with VBAP is close to the isotropic reference up to dip at 3.3 kHz and starts diverging strongly from the reference at about 4.3 kHz, showing the array topology related peak at about 4.7 kHz. Without VBAP (DIR 96, dotted) the reference is approximated more accurately, as can be expected from the spatially uniformly distributed 96 VRS optimized for sphericity.

To maintain clarity, only the selected coherence function estimates for 6, 24, and 96 VRS are shown in Fig 5. For all numbers of VRS, assessment of the coherence curves for 170 mm receiver spacing yielded estimates of the upper frequency limit of correspondence (+- 0.1) with the reference curve of 0.7, 1.4, 1.7, 3, 3.3, 4.3 kHz for 6, 12, 24, 48, 96, 9999 VRS, respectively.

For the BTE microphones (15.6 mm spacing between front and rear, bottom panel of Figure 5), as expected from theory, the closer receiver spacing allows for a considerably better approximation of the reference coherence (+- 0.1) up to at least 5 kHz even for 6 VRS, about 8 kHz for 12 VRS and more than 10 kHz for bigger numbers of VRS. Here, the reference curve only decays to a coherence of about 0.4 at 8 kHz and the first zero crossing occurs at 11 kHz outside the depicted range in contrast to 1 kHz for the 170 mm case in the top panel of Fig 5.

**Receiver rotation**

To investigate the effect of head rotations in VAEs, rotation in the horizontal plane ranging from 0° (as used before) to 358° in 2° steps was assessed for the 170 mm and 15.6 mm spaced receiver pairs for different numbers of VRS. Figure 6 shows the maximum and minimum coherence range of the coherence function estimates as shaded areas outlined by different line styles (see legend). In an ideal isotropic sound field, coherence would be independent of the rotation and follow the reference function (solid black). For the 170 mm spacing (upper panel of Fig 6) the coherences ranges are narrow and follow the reference curve independent of the number of VRS below 700 Hz. Above 700 Hz, the coherence ranges diverge the more, the smaller the number of VRS. Particularly for the lowest number of 6 VRS, a wide coherence range is already observed above 1 kHz reaching values up to 1 for frequencies above 2 kHz. To further assess the variability of the coherence for 6 VRS, two example curves for 0° (replotted from Figure 5) and 60°, as used in the psychoacoustic experiments (condition A, Ar) are shown. For increasing numbers of VRS, the ranges are more centred around the reference and decrease in width. The upper frequency for which the maximum deviation of coherence can be considered reasonably close (+- 0.1) to the reference is about 0.7 kHz for 6 VRS, 1.1 kHz for 12 VRS, 1.5 kHz for 24 VRS, 2.3 kHz for 48 VRS and about 2.5 kHz for 96 VRS.

For the BTE microphones (15.6 mm spacing, bottom panel of Figure 6) the effect of rotation is considerably smaller and all coherence function estimates for more than 6 VRS are close to the reference function (+- 0.1) up to 8 kHz and higher. For 6 VRS, this limit is reached at about 5 kHz.

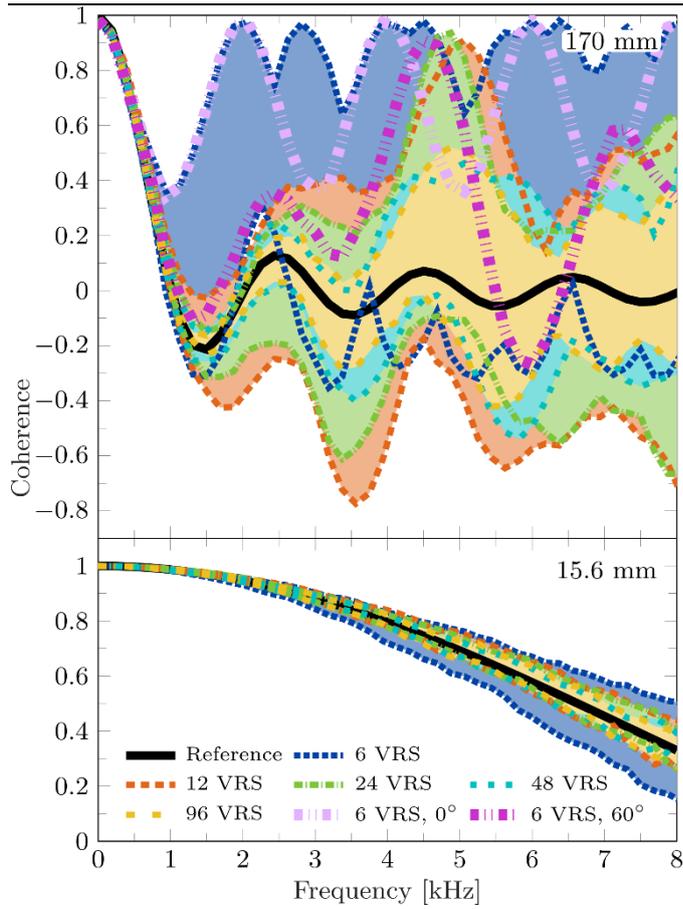

*Figure 6: Coherence ranges for pairs of 170 mm (top panel) and 15.6 mm (bottom panel) spaced omnidirectional receivers throughout a full azimuth rotation in relation to the 86 channel loudspeaker array for different numbers of VRS. For an ideal reproduction of the isotropic sound field, the range would be infinitely narrow and follow the solid black reference curve. As an example, coherence functions for 6 VRS at a receiver orientation of 0° and 60° are illustrated.*

## Comparison to measured BRIRs

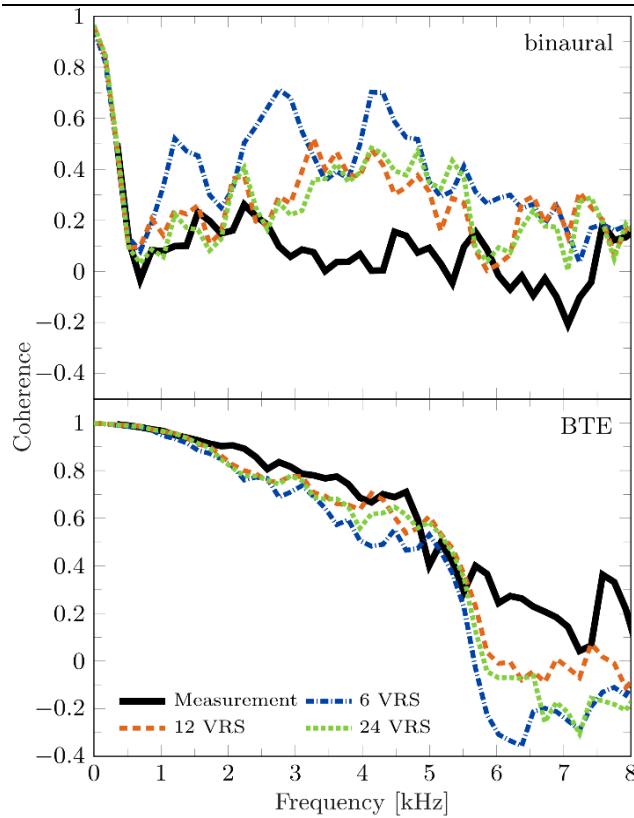

*Figure 7: Coherence function estimates for simulated and measured (solid black line) binaural (top panel) and BTE (bottom panel) impulse responses in the example staircase room.*

As an example of a typical VAE application, the interaural (between eardrums) and BTE coherence for measured BRIRs in the staircase is shown in Figure 7 (solid black line) in comparison to the simulated BRIRs rendered via VBAP in the loudspeaker array. Since the BRIRs also contain direct sound and early reflections, the resulting signals are not a test case for pure diffuse sound field reproduction, but better reflect conditions that might occur during a listening tests or hearing aid evaluation in a VAE. The limited set of 6, 12, and 24 VRS was chosen for increased clarity in the figure and as these numbers of VRS are expected to show the largest effects based on the results so far.

Coherence functions between the two eardrum channels (binaural, upper panel) are conceptually comparable to those of the 170 mm spaced omnidirectional receivers in the evaluation of the isotropic diffuse field. Both simulated and measured curves show a similar behavior up to a frequency of about 700 Hz for 6, 12, and 24 VRS. At higher frequencies, 6 VRS result in a higher coherence than the measured reference as can be expected from the isotropic diffuse field simulation (compare to upper panel of Figure 5). The simulations utilizing 12 VRS start to deviate from the reference at about 700 Hz while 24 VRS follow the reference more closely up to above 1 kHz. At about 2.3 kHz, both simulations show a peak in the coherence which is not present in the measurement and then approximate the measurement again up to 2.6 kHz. Above 2.6 kHz frequencies, the simulated coherence for 12 and 24 VRS is generally higher than in the measurement. The differences between 12 VRS and 24 VRS are generally small and no systematic behavior is apparent. For higher numbers of VRS (not shown here) similar results to 24 VRS are obtained.

For the BTE microphones (15.6 mm spacing, bottom panel of Figure 7), all coherence function estimates independent of VRS number and the measurement are very similar up to about 2 kHz. Up to 5 kHz, the simulations show a reasonably close approximation of the measured coherence, with increasing divergence from the measurement for higher frequencies. Larger errors are observed for 6 VRS compared to 12 and 24 VRS.

## Discussion

*Number of VRS and array topology.* A large effect of number of VRS and rotation, as well as of array topology on coherence was observed for the ear distance (170 mm) spacing: Independent of the concept of VRS for rendering diffuse late reverberation, the inhomogeneous distribution of loudspeakers on the spherical surface of the investigated 86-channel loudspeaker array leads to a decrease in reproduction accuracy for the reproduction of spherical isotropic sound fields above 4 kHz when compared to the spatially more evenly distributed Fibonacci-lattice layout with a similar number of loudspeakers. Often the reproduction of diffuse reverberation is, however, not the only goal. In many listening scenarios, auditory objects tend to be predominantly located in the horizontal plane, where an inhomogeneous layout can minimize rendering artefacts of the direct sound. Also, the reproduction of cylindrical isotropic sound fields (not shown here) benefits from the high spatial resolution in the horizontal plane. The comparison between renderings with and without VBAP show a detrimental effect of the inhomogeneous array topology for larger numbers of VRS for which the sphericity of the spatial VRS distribution is high. Nevertheless, any increase in the number of VRS still leads to an increase in the upper frequency limit for accurate reproduction for the numbers of VRS considered in this study.

*Relation to perception.* In the perceptual findings, 12 or more VRS were indistinguishable from the highest spatial resolution of the reverberation rendering for a majority of listeners. This coincides with an accurate reproduction of sound field coherence up to a frequency of at least 1500 Hz for the 170 mm spaced omnidirectional receivers and a reasonable approximation of coherence in that frequency range in the ear drum channels (staircase example), in line with the upper frequency limit for the relevance of interaural time differences commonly found in literature (e.g., Klumpp and Eady, 1956; Zwislocki and Feldman, 1956; Brughera et al., 2013). Given that listeners are less sensitive to changes in coherence when values are close to 0 (e.g., Robinson and Jeffress, 1962), it is likely that inaccuracies at higher frequencies, where coherence generally drops to low values in the simulations as well as in the reference case, are perceptually less relevant. Similarly, the observed deviations in coherence below 1.5 kHz for more than 12 VRS in the simulations considering rotation (Figure 7) as well as in the test case scenario are well within the more recently derived narrow-band JNDs stated by Walther and Faller (2013).

The strong fluctuations of coherence for rotation of the 170-mm-spaced receiver for low numbers of VRS (see, e.g., 0° and 60° for 6 VRS in the upper panel of Figure 7) likely played a role in the results of the perceptual evaluation for the rotated condition, although the isotropic case in the technical evaluation and the anisotropic case in the perceptual evaluation are not directly comparable, Thus, considering head rotations, the technical evaluation suggests that an increased number of VRS might be beneficial when deviating from a static (optimal) alignment of head orientation and loudspeaker array. An additional parameter that may become relevant in practical applications is the relative orientation of the VRS and the loudspeaker array. However, this parameter was kept fixed by exploiting rotational symmetry of the array geometry and hence kept out of the scope of this study.

Based on both perceptual and technical findings, 12 directions can thus be generally considered sufficient for the spatial rendering of late reverberation. In more critical conditions with strong inhomogeneity in late reverberation, 24 directions are required.

*BTE microphones and hearing aid algorithms.* The isotropic sound field reproduction for the BTE inter-microphone spacing of 15.6 mm showed a close approximation of the reference coherence up to at least about 5 kHz even for 6 VRS. For the staircase test case an accurate reproduction of coherence up to about 4500 Hz can be achieved with 12 or more VRS. This covers the entire frequency range that is most commonly used for hearing aid signal processing. The performance that can be expected from any multi microphone hearing aid signal processing algorithm would have to be evaluated individually, depending on the assumptions about the sound field (also beyond coherence) and error tolerance of that particular algorithm.

In line with the current findings, Grimm et al. (2015) showed that for a horizontal loudspeaker array, the function of a monaural adaptive differential microphone showed no further improvement for more than 8 loudspeakers in the frequency range up to 4 kHz. For a binaural noise reduction algorithm, using the larger ear distance, at least 18 loudspeakers were required, in agreement with the higher number of VRS required in the current study to approximate coherence for the ear distance. Considerably higher numbers of up to 72 loudspeakers were only required in Grimm et al. (2015) for off-center positions in the loudspeaker array, which were not considered here.

The current findings are also in general agreement with faithful performance of binaural hearing aids in a spherical 41-loudspeaker array as observed by Oreinos and Buchholz (2016) which is broadly comparable to the 48 VRS condition for the reproduction of an isotropic sound field. However their loudspeaker driving functions and conditions differ from the current study making a more detailed comparison difficult.

*Application to real-life scenarios and limitations.* The simulation of the highly reverberant staircase example case showed a reasonably accurate reproduction of the interaural coherence for the perceptually relevant frequency range up to 1-1.5 kHz. The similarities in the observations for both the simulated isotropic sound field and the example case suggest that the proportion of energy in the late reverberation in comparison to the energy in the total impinging sound is large enough to play a significant role in the coherence of the receiver signals. With shorter reverberation times, this proportion is expected to be smaller and the relevance of spatial reproduction of late reverberation might be reduced. However, a relaxation of the rendering requirements could not be confirmed by the seemingly room-independent perceptual results for rooms with isotropic reverberation as presented in III.A.

While RAZR was used as specific implementation to find the parameters that determine the minimum number of VRS, the current considerations are universally applicable to other implementations of late reverberation rendering, provided they generate incoherent VRS signals. Similarly, the loudspeaker array geometry that has been the basis for the investigations is not universal, but not atypical in terms of layout and number of loudspeakers and provides an orientation for applications in other arrays.

## Conclusions

For simulating virtual acoustic environments, 3-dimensional diffuse late reverberation is often approximated by a limited number of discrete directions, from which incoherent late reverberation signals are rendered (virtual reverberation sources, VRS). A low number of VRS reduces computational complexity of the underlying simulation and rendering. Considering human perception and physical properties of the reproduced sound field in VAEs, the following conclusions can be drawn:

- Perceptual evaluation shows that the ability of subjects to discriminate a rendering with a lower number of VRS from the reference condition depends on the source material. Deterministic transient stimuli reveal more differences than random speech tokens, which can be assumed to better represent applications using running speech.

- For spherically isotropic late reverberation, 12 spatially distributed VRS are sufficient for rendering without causing perceivable differences to the 96 VRS reference condition, while anisotropic reverberation requires 12 to 24 VRS. An analysis of interaural coherence indicates that these numbers coincide with sufficiently close approximation of coherence at ear distance up to 1.5 kHz or more.

- Variations in interaural coherence caused by rotation of the head relative to the VRS are particularly critical for the lowest number of 6 VRS assessed in the current study. Although such a low number might be acceptable for applications limited to running speech and isotropic reverberation, it can generally not be recommended for applications where the head can be freely moved in relation to the VRS.

- At ear distance, reproduction of sound field coherence is affected by both the arrangement of a particular loudspeaker array and the number of VRS. In the current spherical 86-channel loudspeaker array using VBAP for rendering of the VRS, interaural coherence for the spherically isotropic case was reproduced up to 3.3 kHz for the highest number of 96 VRS and for a real-life test scenario of a simulated staircase room up to at least 1.5 kHz for 12 and more VRS.

- For a typical BTE microphone spacing, coherence can be well reproduced up to at least 5 kHz with 12 VRS for all conditions considered here. For higher numbers of VRS, no further improvements are observed in the considered frequency range up to 8 kHz.

## Acknowledgements

The authors would like to thank Sabine Hochmuth for providing additional OLSA recordings used in the listening test and Florian Denk for support for the HRTF measurements.

## Declaration of Conflicting Interests

## Funding

The authors disclosed receipt of the following financial support for the research, authorship, and/or publication of this article:

This work was funded by the Deutsche Forschungsgemeinschaft (DFG, German Research Foundation) – [Project-ID 352015383] – SFB 1330 C 5